\documentclass{aastex}
\usepackage{apj}
\shorttitle{DIRECT Stellar Catalog. I}
\shortauthors{Macri {\it et al.}}
\def \nd {\nodata}

\def \de {^{\circ}}
\def \hr {^{\rm h}}
\def \mi {^{\rm m}}
\def \se {^{\rm s}}

\begin{document}
\title{The DIRECT Project: Catalogs of Stellar Objects in Nearby Galaxies.
I. The Central Part of M33\altaffilmark{1}}

\author{L.M.~Macri, K.Z.~Stanek\altaffilmark{2}, D.D.~Sasselov\altaffilmark{3}
\& M.~Krockenberger}
\affil{Harvard-Smithsonian Center for Astrophysics, 60 Garden St., Cambridge MA
02138, USA}
\email{lmacri, kstanek, sasselov, krocken@cfa.harvard.edu}

\and

\author{J.~Kaluzny}
\affil{N. Copernicus Astronomical Center, Bartycka 18, PL-00-716 Warszawa,
Poland}
\email{jka@camk.edu.pl}

\altaffiltext{1}{Based on observations collected at the Fred L. Whipple
Observatory 1.2-m telescope and at the Michigan-Dartmouth-MIT 1.3-m telescope.
\baselineskip=18pt}
\altaffiltext{2}{Hubble Fellow}
\altaffiltext{3}{Alfred P. Sloan Foundation Fellow}
\begin{abstract}

The DIRECT project aims to determine direct distances to two important galaxies
in the cosmological distance ladder -- M31 and M33 -- using detached eclipsing
binaries (DEBs) and Cepheids. The search for these variables requires
time-series photometry of large areas of the target galaxies and yields
magnitudes and positions for tens of thousands of stellar objects, which may be
of use to the astronomical community at large.

During the first phase of the project, between September 1996 and October 1997,
we were awarded 95 nights on the F. L. Whipple Observatory 1.2~m telescope and
36 nights on the Michigan-Dartmouth-MIT 1.3~m telescope to search for DEBs and
Cepheids in the M31 and M33 galaxies. This paper, the first in our series of
stellar catalogs, lists the positions, three-color photometry, and variability
indices of 57,581 stars with $14.4 < V < 23.6$ in the central part of M33. The
catalog is available from our FTP site.
\end{abstract}


\section{Introduction}

The DIRECT project \citep{ka98,st98} started in 1996 with the long-term goal of
obtaining distances to two important galaxies in the cosmological distance
ladder -- M31 and M33 -- using detached eclipsing binaries (DEBs) and Cepheids.
These two nearby galaxies are the stepping stones in most of the current effort
to understand the evolving universe at large scales. Not only are they
essential to the calibration of the extragalactic distance scale, but they also
constrain population synthesis models for early galaxy formation and
evolution. However, accurate distances are essential to make these calibrations
free from large systematic uncertainties.

The search for detached eclipsing binaries and Cepheids in our target fields
requires the detection of a large number of stellar objects in our CCD frames
and the repeated measurement of their fluxes over a relatively large time
baseline, usually of the order of 1-2 years. Since the goal of the project
is not simply the detection of these variables but the determination of
accurate distances to the target galaxies, we must also undertake a rigorous
absolute calibration of our photometry. The resulting catalogs of objects 
contain tens of thousands of objects, out of which we only select a few
hundreds for distance-scale work. However, the astronomical community at
large may benefit from the existence of an accurate, well-calibrated list of
objects in these nearby, often-studied galaxies. This is our rationale for
the publication of these series of catalog papers.

Messier 33 (NGC 598) is one of the main components of the Local Group of
galaxies. It is classified as a SA(s)cd galaxy in the Third Reference Catalog
of Galaxies \citet{rc3} and as a Sc(s)II-III in the Revised Shapley-Ames
Catalog \citet{rsa}. It is located at a R.A. of $1\hr 34\mi$ and a Declination
of $30\de 40\mi$ (J2000.0), and it has major and minor $B_{25}$ isophotal
diameters of $71\arcmin$ and $42\arcmin$, respectively. It has been extensively
studied, appearing in more than 1000 publications. One of first was that of
\citet{hu26}, who stated in the abstract of his paper that ``... [i]ts great
angular diameter and high degree of resolution, suggesting that it is one of
the nearest objects of its kind, offer exceptional opportunities for detailed
investigation.''

The present work will describe the details of the observations (\S2), the
reduction and absolute calibration of the data (\S3), the creation of the
stellar catalog (\S4) and the results of our consistency checks (\S5) for
three CCD fields in the central part of M33. The analysis of the variable stars
located in these fields will be analyzed in two upcoming papers by \citet{ma00}
and \citet{st01}.

\section{Observations}

Our observations of the central region of M33 were primarily carried out at the
Fred L. Whipple Observatory (hereafter FLWO) 1.2-m telescope.  We used
``AndyCam'' \citep{sz00}, a thinned, back-illuminated, AR-coated Loral $2048^2$
pixel CCD camera with a plate scale of $0.317\arcsec/$pixel, or an effective
field of view of $10\farcm 8$. The filters used during our program were
standard Johnson $B$ and $V$ and Cousins $I$. Additional $I$-band data were
collected at the Michigan-Dartmouth-MIT Observatory 1.3-m McGraw-Hill
telescope. We used ``Wilbur'' \citep{me93}, a thick, front-illuminated Loral
$2048^2$ pixel CCD camera. The plate scale and field of view were almost
identical to that of ``AndyCam.''

We observed three fields located north, south and south-west of the center of
M33, which we labeled M33A, B and C. The J2000.0 center coordinates of the
fields are: M33A, R.A. = $01\hr 34\mi 05.1\se$, Dec.= $30\de 43' 43''$; M33B,
R.A. = $01\hr 33\mi 55.9\se$, Dec.= $30\de 34' 04''$; M33C, R.A. = $01\hr 33\mi
16.0\se$, Dec.= $30\de 35' 15''$. Figure 1 shows the boundaries of these fields
overlaid on a digitized image of the galaxy from the POSS-I survey\footnote{
The Digitized Sky Surveys were produced at the Space Telescope Science
Institute under U.S. Government grant NAGW-2166. The National Geographic
Society -- Palomar Observatory Sky Atlas (POSS-I) was made by the California
Institute of Technology with grants from the National Geographic Society.},
while Figure 2 shows a mosaic of the survey fields, created with our CCD
data. At FLWO, we obtained $V$ and $I$ data on 42 nights and $B$ data on 13
nights. At MDM, we obtained $I$ data on 10 nights.  Exposure times were 1200s
in $B$, 900s in $V$ and 600s in $I$. Fields were observed repeatedly on each
night in $V$ and $I$, so the actual number of exposures per field in those
filters is around 110 and 60, respectively. Standard star fields from
\citet{la92} were observed on one photometric night. Table 1 presents a log of
our observations.

\section{Data reduction and calibration}

\subsection{PSF photometry}

Paper I of the DIRECT variable star series \citep{ka98} contains a detailed
description of the data reduction and PSF photometry. Only a brief summary of
these procedures is presented here. The CCD frames were processed using
standard CCDPROC routines under IRAF\footnote{IRAF is distributed by the
National Optical Astronomy Observatories, which are operated by the
Associations of Universities for Research in Astronomy, Inc., under cooperative
agreement with the NSF.}. Photometry was obtained using the DAOPHOT and ALLSTAR
programs \citep{st87,st92}, using a Tcl/Tk-based automated reduction pipeline.

Point-spread functions (PSFs) were calculated from bright and isolated stars
present in each frame, following an iterative process. Figure 3 shows a
histogram of the seeing for the three filters; median FWHM values were
$1.5\arcsec$ for I and $1.8\arcsec$ for B and V. After running DAOPHOT and
ALLSTAR on all frames, we selected an image of particularly good quality (in
terms of seeing and depth) as a ``template'' frame. ALLSTAR was run again in
``fixed-position'' mode on all other images, using the transformed object list
from the template frame as input. The resulting photometry lists were
transformed back into the coordinate and instrumental magnitude system of the
template image. The latter was accomplished by computing a local magnitude
offset for each star, using high SNR stars ($\sigma < 0.03$~mag) located within
a radius of 350 pixels. In cases where few stars met these conditions, the
search radius was increased to 750 pixels, or a global median offset was used
as a last resort. The magnitude offset between each frame and the template
image was recorded in a log file for future use (see below). The typical
uncertainty in this offset was 0.02~mag.

Thus, for each field and filter combination, the output of our automated
reduction pipeline consisted of one ALLSTAR file for each frame, with positions
and PSF magnitudes in the coordinate and photometric systems of its template
frame. The ALLSTAR files pertaining to a particular field and filter
combination were matched and merged, to arrive at nine final photometry
databases (3 fields $\times$ 3 filters).

The instrumental PSF magnitudes present in the databases had to transformed
into the standard system. This procedure can be separated into three steps: i)
transform PSF magnitudes in the instrumental system of the template frame to
PSF magnitudes in the instrumental system of the photometric frame; ii)
transform PSF magnitudes in the instrumental system of the photometric frame to
aperture magnitudes in the instrumental system of the photometric frame; iii)
transform the instrumental system of the photometric frame to the standard
system. These steps are described in detail below.

\subsection{Aperture corrections}

The first step of the photometric calibration process was the transformation of
the PSF magnitudes of the photometry database from the magnitude scale of the
template frame to the magnitude scale of another frame, taken under photometric
conditions (hereafter referred to as the ``photometric frame''). This was
easily achieved by applying a magnitude offset of equal size and opposite sign
to the one which had already been determined (as part of our automated
pipeline) to exist between the template frame and the photometric frame.

The second step of the process was the transformation of PSF magnitudes into
aperture magnitudes, through the determination of aperture correction
coefficients. Given the crowded nature of our fields, their rapidly-varying sky
backgrounds, and the relatively poor seeing of our photometric night, a
thorough approach was required. We chose one frame for each field
and filter from the photometric night, and used the master star lists and the
PSFs derived by our automated pipeline to remove all objects present in these
images, with the exception of bright, isolated stars. Aperture photometry was
carried out on these star-subtracted frames at a variety of radii (ranging
from 10 to 20 pixels, or 3 to 6$\arcsec$). The local sky was characterized
using an annulus extending from 30 to 40 pixels.

The aperture photometry measurements of all bright stars in a particular frame
were examined simultaneously by visually inspecting their curves of growth
(i.e., plots of aperture magnitude versus radius). Objects with unusual growth
curves were discarded. The aperture photometry measurements of the remaining
bright stars (hereafter, ``input stars'') were analyzed using DAOGROW
\citep{ste90}. This program performs an analytical fit to the growth curves of
all input stars in all frames, and the resulting function is used to determine
a mean growth curve for each frame. DAOGROW then uses the best combination of
aperture photometry and growth curve for each input star to calculate its
aperture magnitude at the outermost radius (in our case, 20 pixels). Lastly,
the PSF and aperture magnitudes of all input stars in each frame are used to
derive a mean value of the aperture correction, which is applied to all
objects. The aperture correction coefficients derived using this procedure
ranged from $-0.10$ to $+0.24$~mag, with typical uncertainties of 0.03~mag.

\subsection{Photometric solutions}

Once the instrumental PSF magnitudes in each of the nine databases were
converted to instrumental aperture magnitudes, the last step required to
transform them into standard magnitudes was the derivation of photometric
zeropoints. On 1997 October 9, a photometric night of average seeing quality
for our program (I: $1.8\arcsec$; B and V: $2.0\arcsec$), we observed six
fields from \citet{la92}, containing a total of forty-three standard stars, at
airmasses ranging from 1.12 to 2.12. We performed photometry on the standard
stars using DAOPHOT with the same settings used for the program stars, namely
an aperture radius of 20 pixels and a sky annulus extending from 30 to 40
pixels. We used the IRAF PHOTCAL routines to solve for a photometric solution
of the form

\vskip -8pt
\begin{equation}
M_{std,i} = m_{obs,i} + \chi_i - k'_i X + \xi_{ij} (M_{std,i} - M_{std,j})
\end{equation}
\vskip 3pt

\noindent{where $M_{std,i}$ and $M_{std,j}$ are the magnitudes of a star in
the standard system in the $i$ and $J$ filters, while $m_{obs,i}$ is the
instrumental magnitudes of the same star in the $i$ filter.  $\chi_i$ is the
magnitude zeropoint at $X=0$, $k'_i$ is the airmass coefficient for the $i$
filter, and $\xi_{ij}$ is the color term. The V-band solution was calculated
using both B-V and V-I for the color term; the latter one was used by default
in the calibration process, unless only B and V data were available for a
particular object. The B-band solution was calculated using B-V for the color
term, while the I-band solution was calculated using V-I for the color
term. The values and uncertainties of the coefficients of each term are
presented in Table 2; based on those numbers, we estimate a total uncertainty
of $\pm 0.02$~mag in our solutions.}

Based on the uncertainties associated with PSF magnitude offsets
($\pm0.02$~mag, \S3.1), aperture correction coefficients ($\pm0.03$~mag,
\S3.2) and photometric solutions ($\pm0.02$~mag, previous paragraph), we
estimate a total random uncertainty in our photometric zeropoints of
$\pm0.04$~mag.

\section{The star catalog}

Once the photometric calibrations were applied, we merged the $BVI$ databases
of each field into a single catalogs. Objects were matched from the master B, V
and I star lists of each field and were kept only if they had been detected in
the V band and in either of the B or I bands. Next, we transformed the object
coordinates into the FK5 system using stars from the USNO-A2.0 catalog
\citep{mo98}. We solved for a cubic-order transformation using software
developed by \citet{min99}. The solutions used 30-70 stars and had {\it rms}
values of $0.4\arcsec$.

Lastly, the catalogs of the three fields were merged into a single, master
catalog.  There was --by design-- significant overlap between fields A and B as
well as between fields C and B; objects in these regions were matched to test
the internal consistency of our astrometric and photometric calibrations (see
\S5). To avoid duplication of these objects, we only kept the entry from 
field B.

As described in \citet{ka98}, the magnitude uncertainties reported by
DAOPHOT/ALLSTAR are under-estimated for bright stars and over-estimated for
faint ones. The errors were re-scaled following the precepts established in
that paper. Lastly, we calculated mean BVI magnitudes and V-band $J_S$
variability indices \citep{ste96}. The catalog is presented in Table 3; it
lists IDs, celestial coordinates, mean B, V and I magnitudes and
uncertainties, and $J_S$ indices for 57,581 stars present in our fields.
The catalog can also be retrieved from the DIRECT FTP site at {\tt http://cfa-www.harvard.edu/$\sim$kstanek/DIRECT}.

Figure 4 shows the differential luminosity function of the objects in our
catalog. The turnovers in these luminosity functions indicate incompleteness
below $\sim22$~mag for B and V, and $\sim20$~mag for I. Figure 5 shows
color-magnitude diagrams of our catalog stars. A faint plume of foreground
stars from our own Galaxy can be seen in the region $0.4 < B-V < 1.2$, $V <
20$. The feature is substantially diminished relative to the one seen in the
CMD of M31 in \citet{ka98} due to the difference in galactic latitude between
these two objects ($l\sim -22\de$ for M31 and $l\sim -31\de$ for M33).

We flagged objects as candidate variables if they met two requirements:
a $J_S$ index larger than 0.75, and a V-band magnitude uncertainty larger
than 0.04~mag. The second criterion was introduced to remove bright stars
with small variability from our sample of candidate variables (in this
data set, it removed 107 stars with $V < 19.5$~mag). Our final sample
of candidate variables consists of 1,298 stars. The panels of Figure 6 show
some global properties of the variable stars present in our catalog.

\section{Test of photometric and astrometric calibrations}

We used $\sim 5000$ objects present in the overlap regions between the survey
fields to check our astrometric and photometric calibration procedures. We
compared the celestial coordinates of these objects and found small offsets
between fields of the order of $0\farcs4 - 0\farcs7$, which are consistent
with the {\it rms} residuals of the astrometric solutions.

We performed an internal test of our photometric calibration by comparing the
mean B, V and I magnitudes of bright stars present in the overlap regions.
We imposed magnitude cuts of 19.5, 19.5 and 19.0~mag in B, V, and I,
respectively, which restricted the number of matches to about 200, 300 and
400, respectively. On average, the offsets were $<0.01$~mag. This indicates
that PSF variations across the field were properly taken into account by
DAOPHOT and our pipeline, and that the aperture corrections were properly
determined. Table 4 lists the values of the offsets and their standard
deviations; Figure 7 shows plots of these comparisons.

We peformed two external tests of our photometric calibration. In the first
test, we matched about 200 objects in common between our Field C and Field 4
of \citet{wfm90}. We compared the mean B, V and I magnitudes of stars brighter
than 20.0~mag in each of the filters (about 25 stars/filter) and found offsets
of the order of $-0.03$~mag (brighter DIRECT magnitudes). In the second
external test of our photometric calibration, we matched about 4000 objects in
common between our Field A and one of the fields of \citet {be01}. We compared
the mean B and V magnitudes of stars brighter than 18.5~mag (about 35
stars/filter) and again found offsets of the order of $-0.03$~mag (brighter
DIRECT magnitudes). Table 4 lists the results of these comparisons, which
are also plotted on Figures 8 and 9.

\section{Artificial star tests}

The differences between our photometry and the \citet{wfm90} and \citet{be01}
photometry are small but consistent. Furthermore, both groups used larger
telescopes (CFHT and WIYN, respectively) under significantly better seeing
conditions that us. Therefore, we decided to undertake artificial star tests
to quantify the level of photometric bias that could arise due to the poorer
spatial resolution of our images.

We used DAOPHOT to inject 2,500 artificial stars into the nine master frames,
using the PSFs previosly derived by our automated reduction pipeline and
taking into account photon noise and other detector characteristics. We
analyzed the frames using the same procedures as in the automated pipeline.
The results were quite similar for the three frames pertaining to each band,
and thus the data files were merged to improve the statistics.  Our results
are presented in Table 5 and in Figure 10.

Bright stars ($15<m<18$) are affected by crowding at the $0.01-0.04$~mag
level. The bias becomes stronger for fainter objects ($m>18$), reaching
$0.05-0.08$~mag. At a given magnitude, the bias increases from B to V to I.
In all cases, the offset induced by crowding is in the same direction as the
offset found between our data and other catalogs.

\section{Summary}

We have observed three fields in the central part of M33 at the Fred
L. Whipple Observatory 1.2-m and the Michigan-Dartmouth-MIT Observatory 1.3-m
telescopes. We have performed PSF photometry of objects in these fields,
calibrated in the standard system with a zeropoint accuracy of $\pm 0.04$~mag.

We have compiled a catalog of positions, B, V and I magnitudes, and V-band
variability indices for 57,581 stars with $14.4 < V< 23.6$. The catalog is
available from our FTP site.

The analysis of the variable star content of these fields will be presented in
two upcoming papers by \citet{ma00} and \citet{st01}.

We would like to thank the telescope allocation committees of the FLWO and
MDM Observatories for the generous amounts of telescope time devoted to this
project. We would also like to thank Peter Stetson, for his photometry
software; Doug Mink, for help with the astrometry; and David Bersier, for
providing us his photometry in advance of publication. LMM would like to thank
John Huchra for his support and comments. KZS was supported by a Hubble
Fellowship grant HF-01124.01-99A from the Space Telescope Science Institute,
which is operated by the Association of Universities for Research in
Astronomy, Inc., under NASA contract NAS5-26555. DDS acknowledges support
from the Alfred P.  Sloan Foundation and from NSF grant No. AST-9970812. JK
was supported by KBN grant 2P03D003.17

\clearpage

\begin{figure}
\plotone{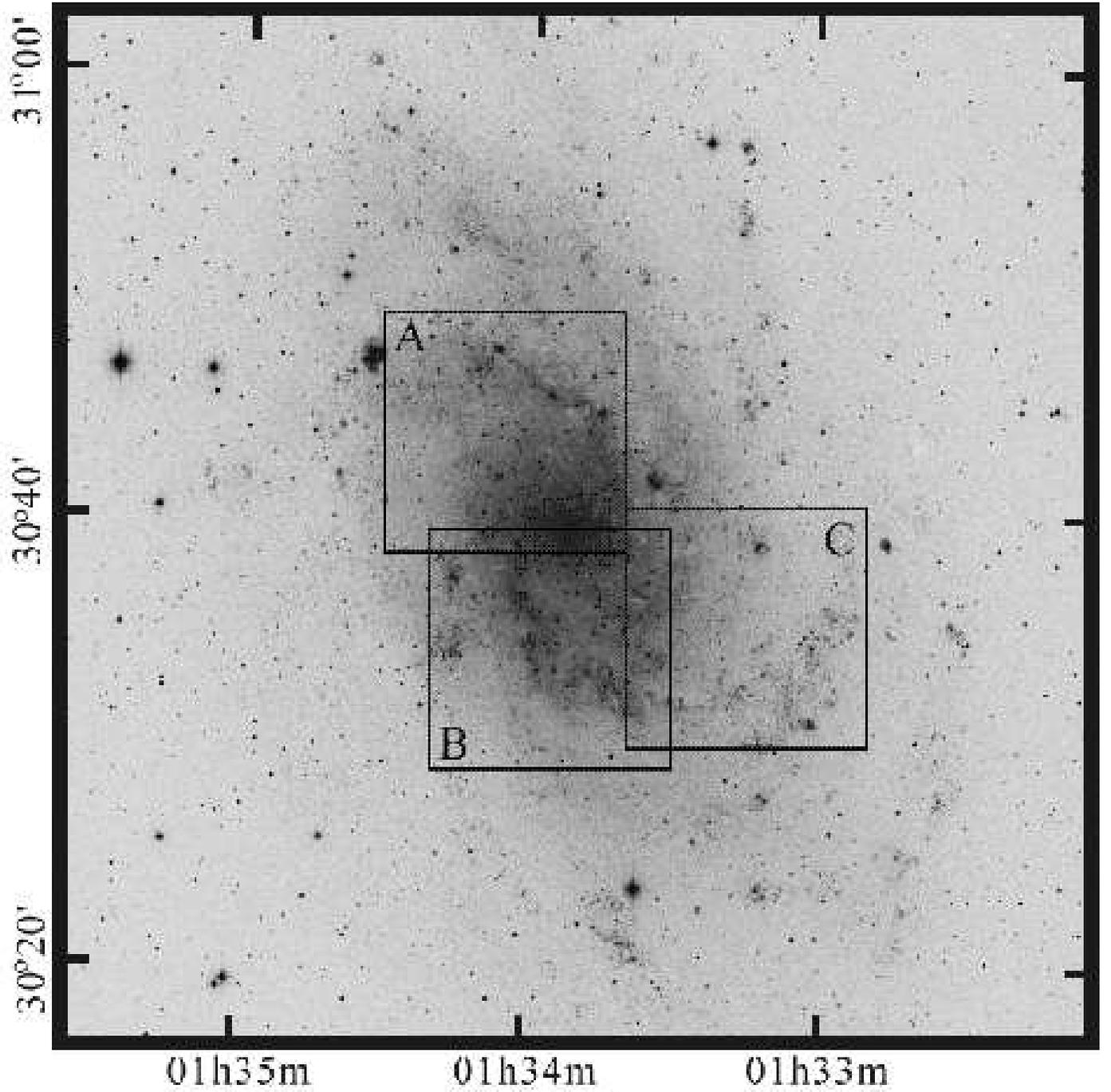}
\caption{Palomar Observatory Sky Survey image of M33, showing the size and
location of fields A-C. Each box is approximately 10$\farcm$8 on a side. North
is up and East is to the left.}
\end{figure}

\clearpage

\begin{figure}
\plotone{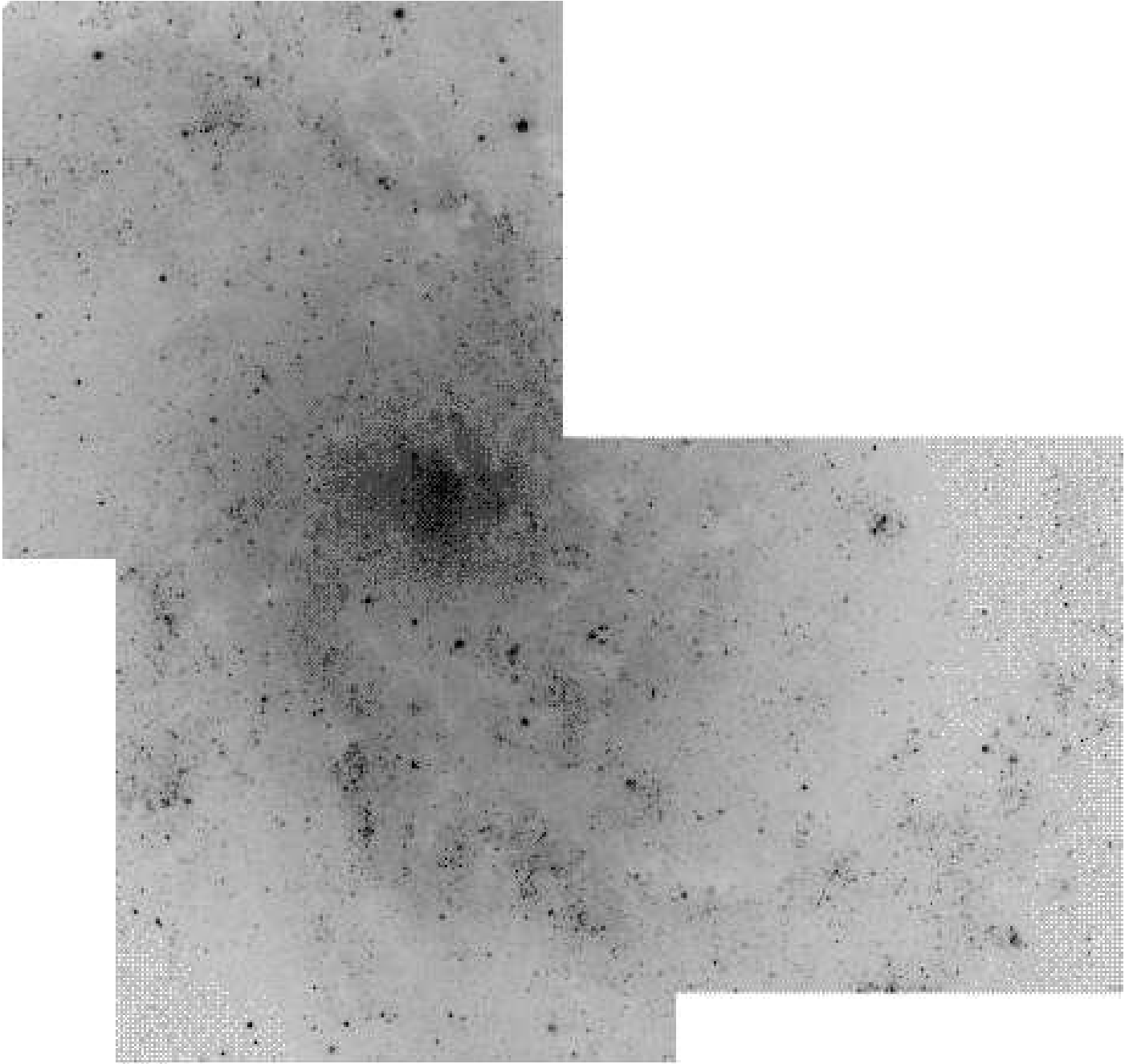}
\caption{Mosaic of the central part of M33, created from CCD images of
our fields.}
\end{figure}

\clearpage

\begin{figure}
\plotone{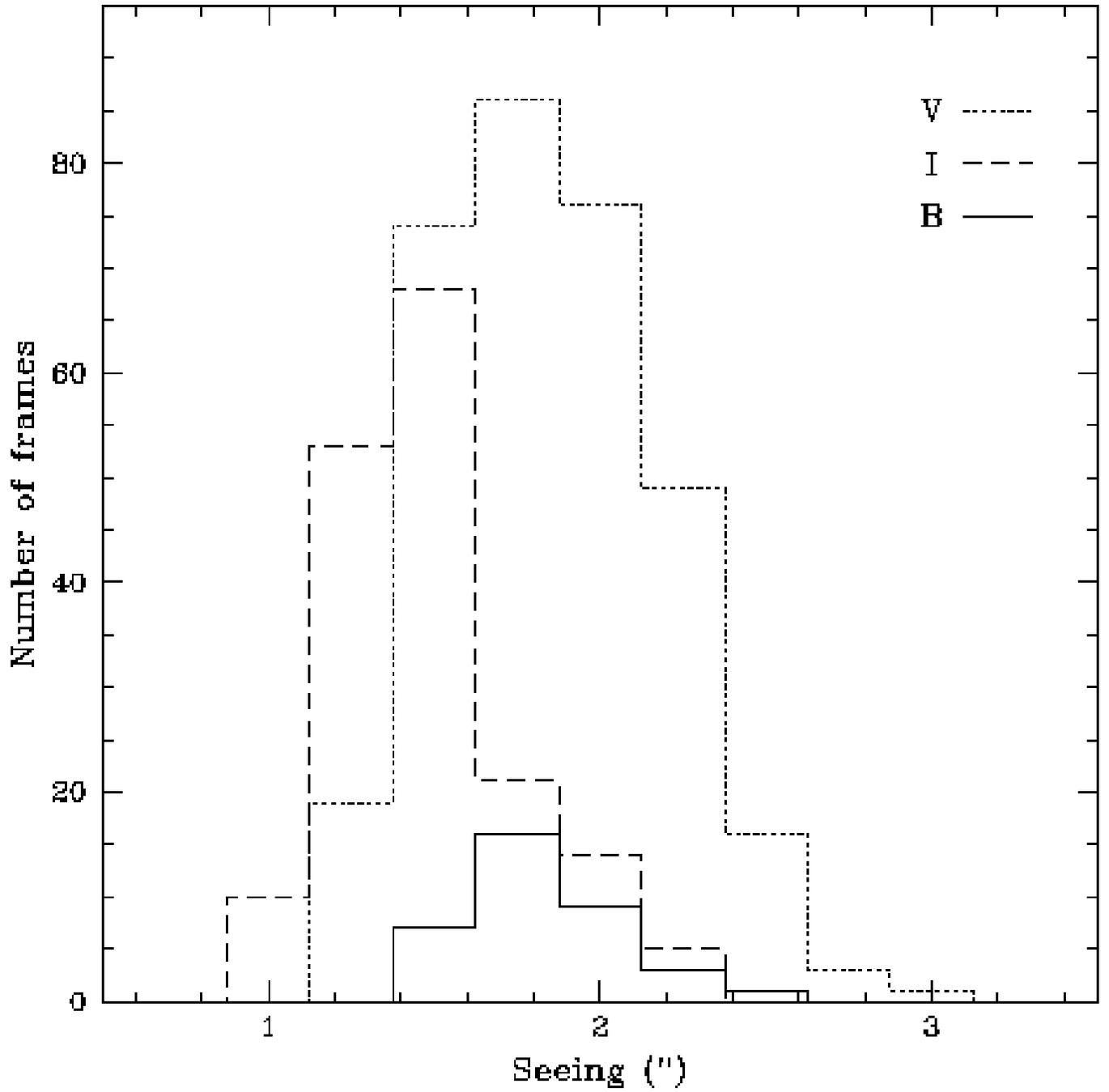}
\caption{Histogram of seeing values for the frames acquired for this project.
Solid, dotted and dashed lines represent the B, V and I band histograms,
respectively.}
\end{figure}

\clearpage

\begin{figure}
\plotone{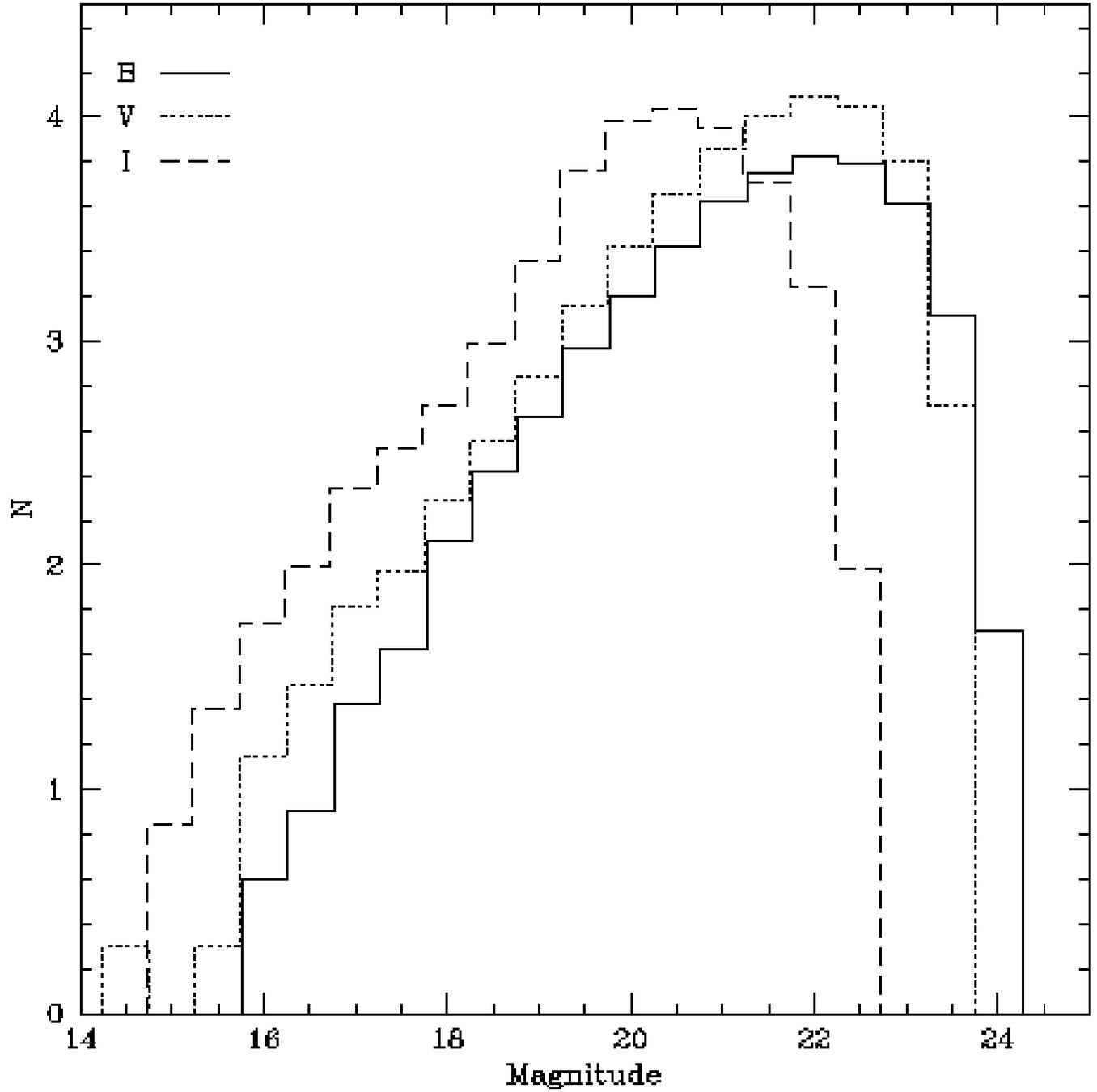}
\caption{Differential luminosity functions for the stars present in our
catalog, for the B (solid), V (dashed) and I (dotted) bands. Our completeness
limits are $\sim22$~mag for B and V and $\sim20$~mag for I.}
\end{figure}

\clearpage

\begin{figure}
\plotone{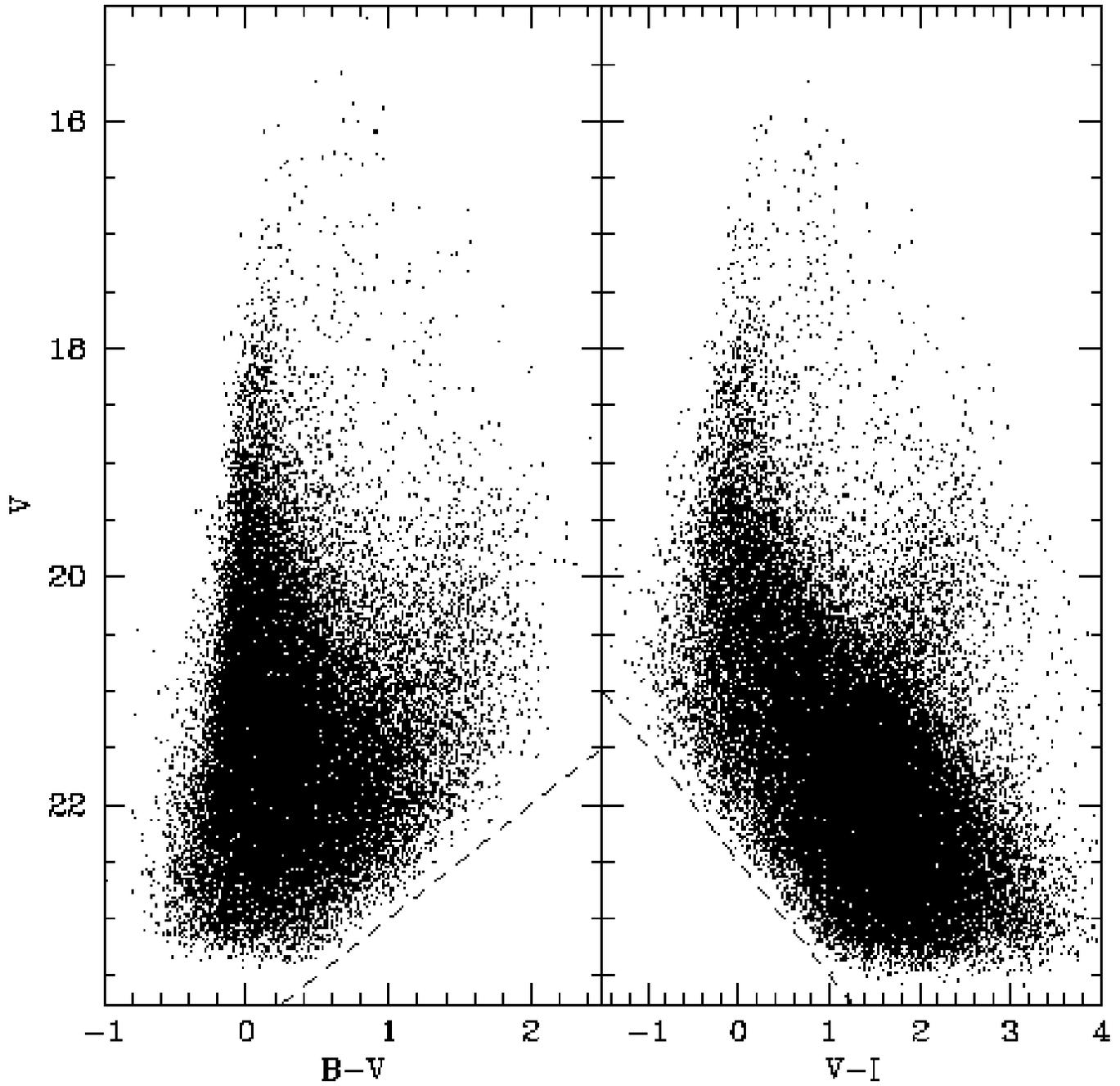}
\caption{Color-magnitude diagrams for the stars present in our catalog.  The
dashed lines indicate the extent of our data, set by our limiting magnitudes
of $B\sim 24$ and $I\sim 22$.}
\end{figure}

\clearpage

\begin{figure}
\epsscale{0.9}
\plotone{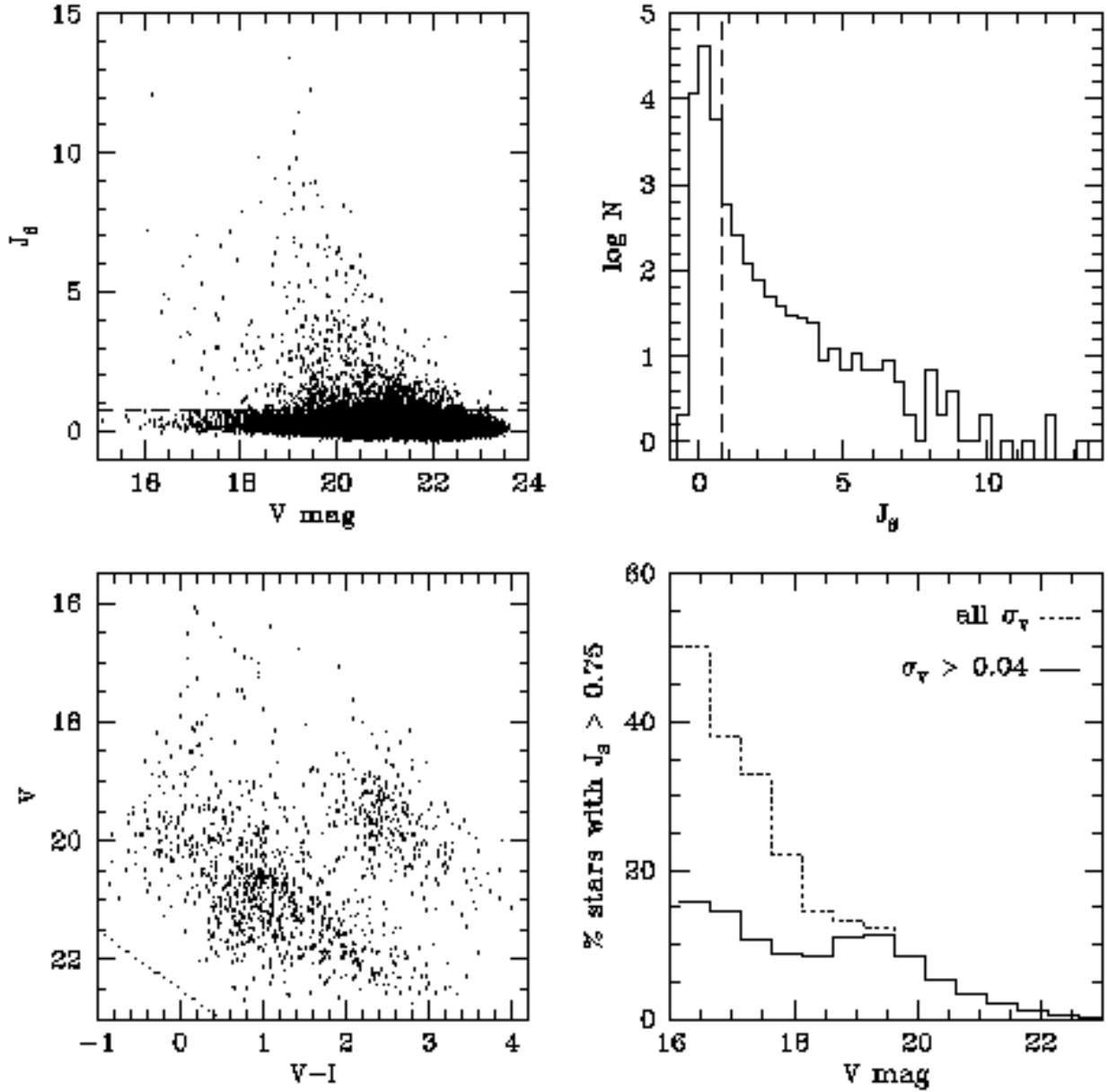}
\caption{Global properties of candidate variables present in our catalog.  Top
left: Distribution of $J_S$ with V mag. The dashed line indicates our threshold
of $J_S=0.75$. Top right: Number of stars in the catalog as a function of $J_S$
value. The dashed line indicates our threshold of $J_S=0.75$.  Bottom left:
Color-magnitude diagram of candidate variables. The dotted line indicates the
extent of our data. Bottom right: Effect of imposing a $\sigma_V > 0.04$~mag
cut in our definition of variability; a large percentage of bright stars are
dropped from the candidate variable sample.}
\end{figure}

\clearpage

\begin{figure}
\epsscale{0.95}
\plotone{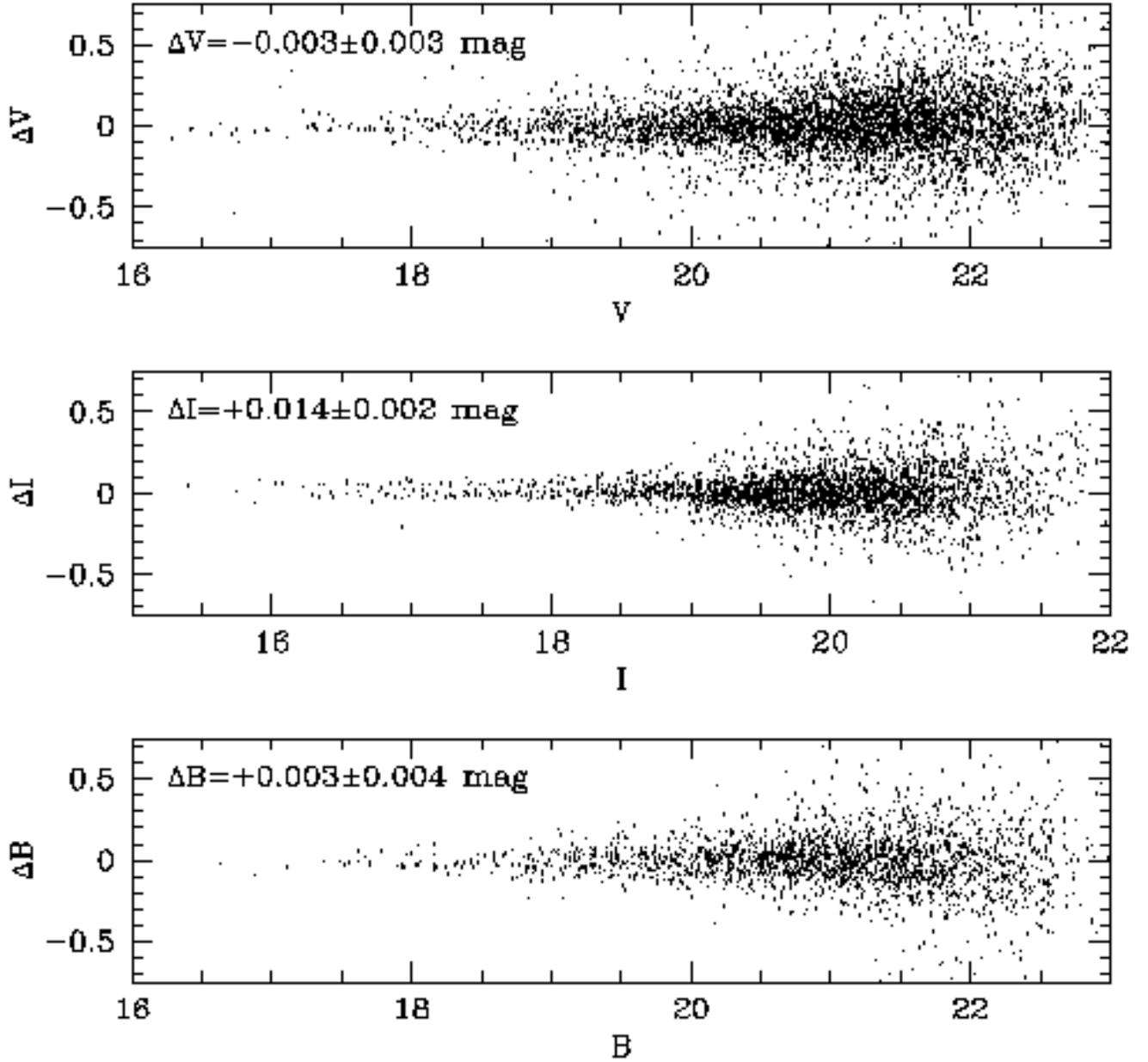}
\caption{Comparison of mean magnitudes for bright stars ($B <
19.5$~mag; $V < 19.5$~mag; $I < 19.0$~mag) located in the overlap
regions between fields A-B and C-B. The photometric zeropoints and
aperture correction coefficients are determined independently for each
field, so these comparisons allow us to check the internal consistency
of our reductions.  The average values and {\it r.m.s.} deviations of the
offsets are listed in the top-left corner of each panel and in Table
4.}
\end{figure}

\clearpage

\begin{figure}
\plotone{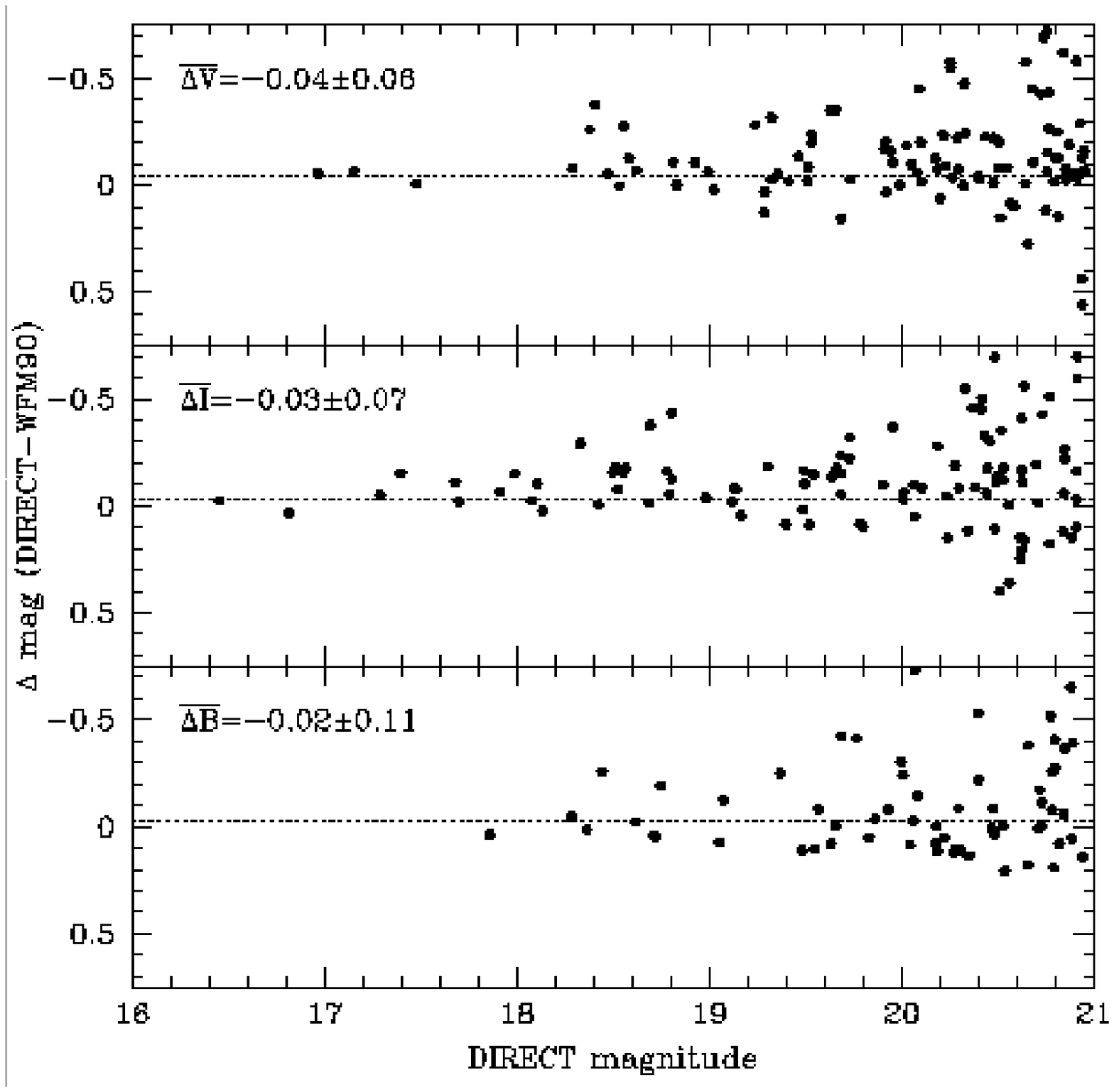}
\caption{Comparison of mean magnitudes for bright stars ($B,V,I <
20$~mag) in common between Field C and Field 4 of \citet{wfm90}. The
average values and {\it r.m.s.} deviations of the offsets are listed in the
top-left corner of each panel and in Table 4.}
\end{figure}

\clearpage

\begin{figure}
\plotone{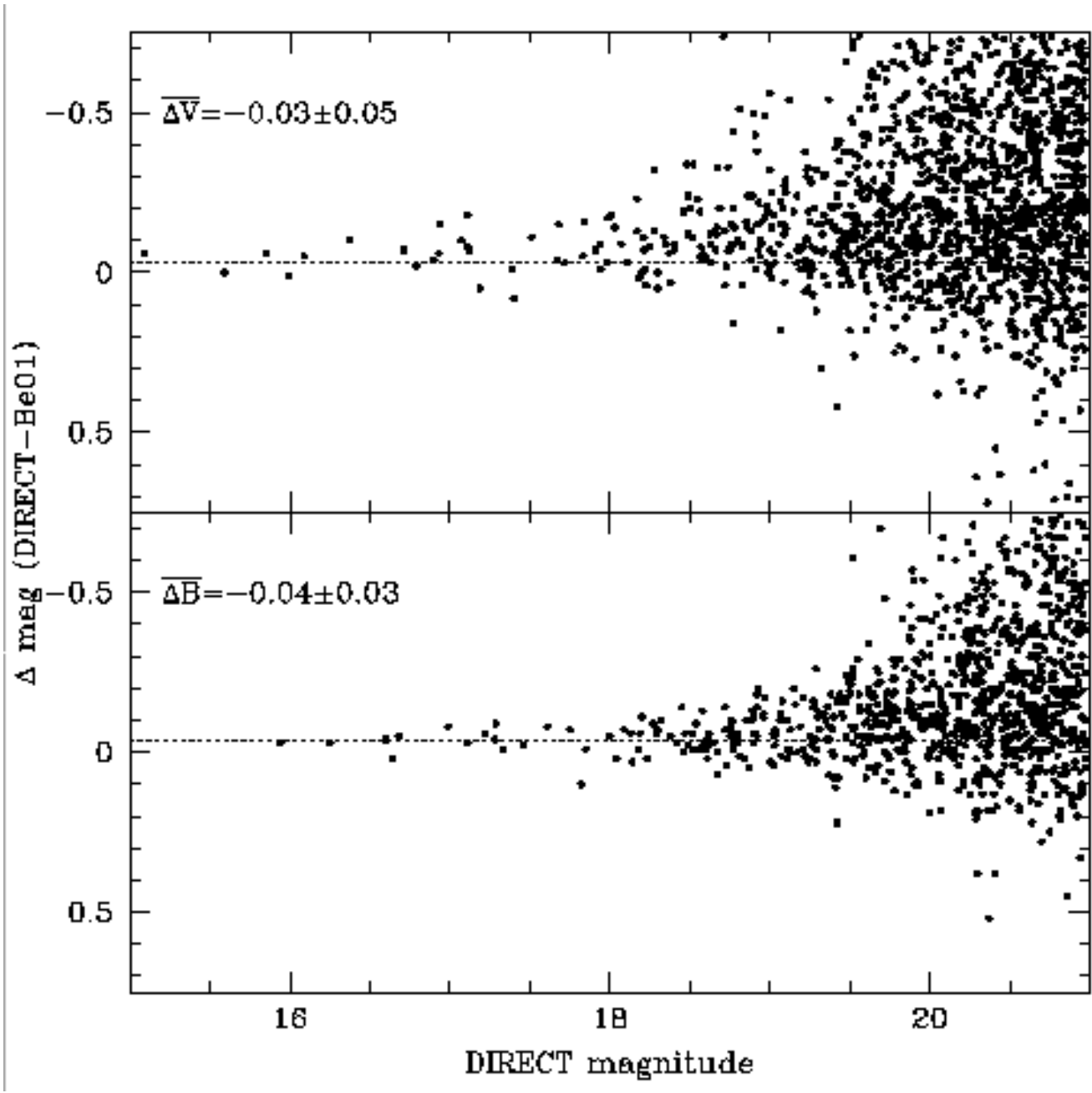}
\caption{Comparison of mean magnitudes for bright stars ($B,V <
18.5$~mag) in common between Field A and one of the fields of
\citet{be01}. The average values and {\it r.m.s.} deviations of the offsets
are listed in the top-left corner of each panel and in Table 4.}
\end{figure}

\clearpage

\begin{figure}
\plotone{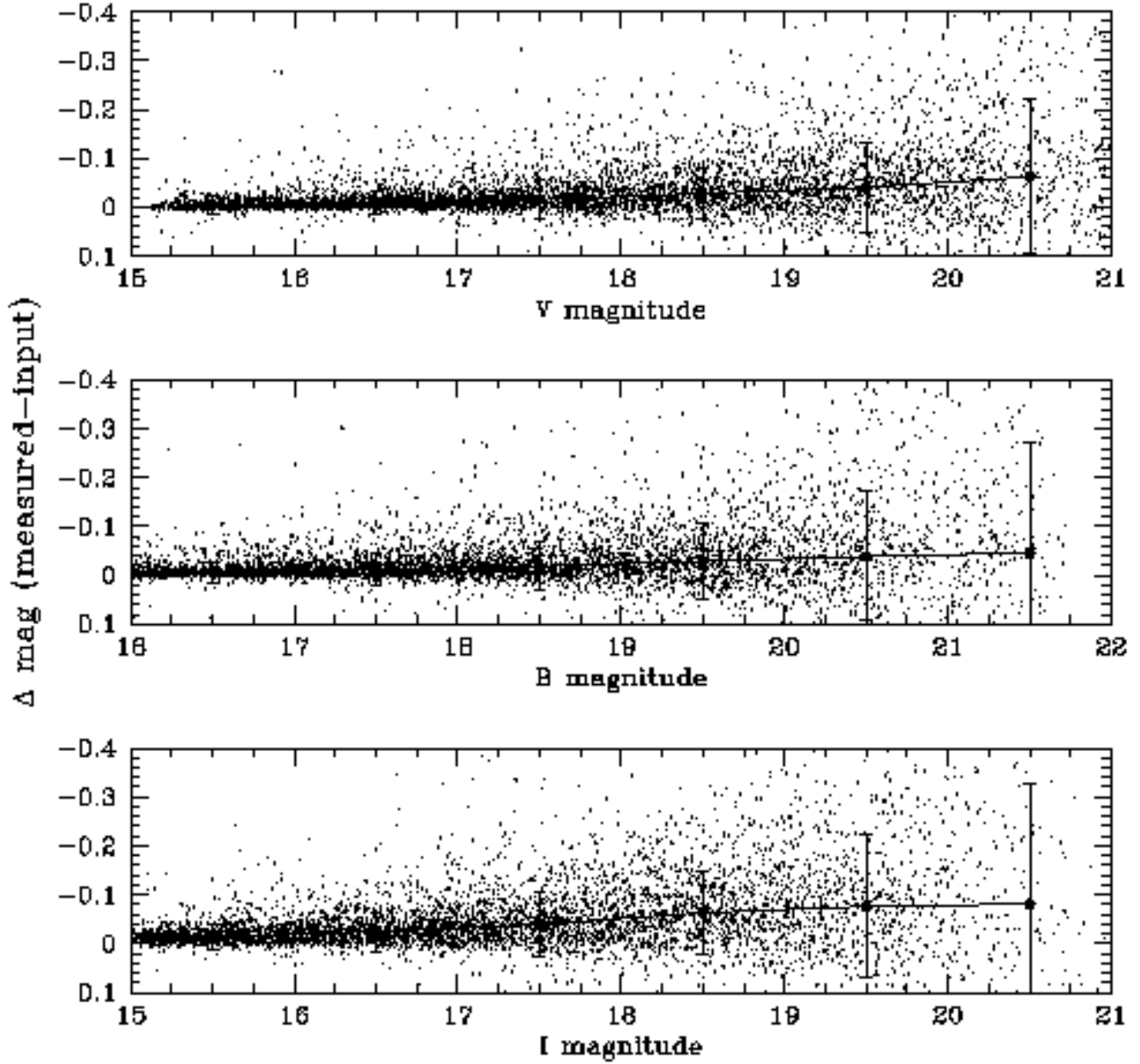}
\caption{Results of the artificial star tests. Individual stars are plotted
using small dots, while median values for one-magnitude intervals are indicated
with solid circles. They are also listed in Table 5.}
\end{figure}

\clearpage

\vskip -1in
\begin{deluxetable}{llcccr}
\tabletypesize{\footnotesize}
\baselineskip=9pt
\tablecolumns{6}
\tablewidth{0pt}
\tablenum{1}
\tablecaption{Log of Observations}
\tablehead{\colhead{UT Date} & \colhead{MJD} & \colhead{Field} & \colhead{Band}
& \colhead{Tel.} & \colhead{Seeing}}
\startdata
1996/09/03 & 329.7615 & M33A & I & F & 1.40 \\
           & 329.7786 & M33A & I & F & 1.24 \\
1996/09/06 & 332.7220 & M33A & V & F & 2.18 \\
           & 332.7504 & M33A & V & F & 2.25 \\
           & 332.7735 & M33A & V & F & 1.77 \\
           & 332.7803 & M33A & V & F & 1.84 \\
           & 332.8479 & M33B & B & F & 2.03 \\
           & 332.8697 & M33B & V & F & 1.82 \\
           & 332.8814 & M33B & V & F & 1.37 \\
           & 332.8932 & M33B & V & F & 1.43 \\
           & 332.9114 & M33C & V & F & 1.39 \\
           & 332.9182 & M33C & V & F & 1.57 \\
           & 332.9434 & M33C & V & F & 1.82 \\
           & 332.9612 & M33C & I & F & 1.28 \\
           & 332.9669 & M33B & I & F & 1.36 \\
           & 332.9753 & M33A & I & F & 1.28 \\
           & 332.9873 & M33A & V & F & 1.34 \\
1996/09/07 & 333.8863 & M33A & V & F & 1.68 \\
           & 333.8930 & M33A & V & F & 1.70 \\
           & 333.9047 & M33A & I & F & 1.07 \\
           & 333.9130 & M33B & I & F & 1.22 \\
\multicolumn{6}{l}{\it Continues in electronic form}\\
\enddata
\tablecomments{Telescope code: F=FLWO; M=MDM.\\
$\dagger$: Photometric night -- Standards observed.}
\end{deluxetable}

\begin{deluxetable}{llllrl}
\tabletypesize{\normalsize}
\baselineskip=12pt
\tablecolumns{7}
\tablewidth{0pt}
\tablenum{2}
\tablecaption{Photometric solution for 1997 Oct 09}
\tablehead{\multicolumn{2}{c}{Filter} & \colhead{$\chi$} &
\colhead{$k'$} & \colhead{$\xi$} & \colhead{\it rms}}
\startdata
B & (B-V) & $-22.953\pm0.025$ & $0.212\pm0.017$ & $-0.033\pm0.011$ & 0.030 \\
V & (B-V) & $-22.714\pm0.014$ & $0.123\pm0.009$ & $ 0.035\pm0.006$ & 0.016 \\
V & (V-I) & $-22.720\pm0.013$ & $0.127\pm0.009$ & $ 0.032\pm0.005$ & 0.016 \\
I & (V-I) & $-22.719\pm0.016$ & $0.064\pm0.010$ & $-0.051\pm0.007$ & 0.021 \\
\enddata
\end{deluxetable}

\begin{deluxetable}{lccccccccccccr}
\tabletypesize{\scriptsize}
\tablecolumns{14}
\tablewidth{0pt}
\tablenum{3}
\tablecaption{Catalog of stars in the central part of M33}
\tablehead{\colhead{ID} & \multicolumn{3}{c}{R.A.} & \multicolumn{3}{c}{Dec.} & \colhead{V} &\colhead{I} &\colhead{B} &
\colhead{$\sigma_V$} & \colhead{$\sigma_I$} & \colhead{$\sigma_B$} & \colhead{$J_S$}}
\startdata
D33 J013251.1+303923.7 & 01 & 32 & 51.11 & 30 & 39 & 23.65 & 19.97 & 18.14 &  \nd  & 0.03 & 0.03 &  \nd &  0.12\\
D33 J013251.1+303741.8 & 01 & 32 & 51.13 & 30 & 37 & 41.81 & 21.61 & 21.65 &  \nd  & 0.11 & 0.28 &  \nd &  0.12\\
D33 J013251.1+303954.9 & 01 & 32 & 51.14 & 30 & 39 & 54.86 & 21.72 & 19.50 &  \nd  & 0.12 & 0.13 &  \nd &  0.06\\
D33 J013251.2+303736.4 & 01 & 32 & 51.17 & 30 & 37 & 36.44 & 20.20 & 19.61 &  \nd  & 0.04 & 0.08 &  \nd &  0.09\\
D33 J013251.2+303907.1 & 01 & 32 & 51.20 & 30 & 39 & 07.09 & 22.87 & 21.82 &  \nd  & 0.26 & 0.27 &  \nd & -0.02\\
D33 J013251.2+303648.7 & 01 & 32 & 51.20 & 30 & 36 & 48.67 & 21.32 & 21.57 &  \nd  & 0.10 & 0.17 &  \nd &  0.04\\
D33 J013251.2+303959.6 & 01 & 32 & 51.22 & 30 & 39 & 59.58 & 22.64 & 21.32 &  \nd  & 0.52 & 0.28 &  \nd &  0.64\\
D33 J013251.2+303607.0 & 01 & 32 & 51.22 & 30 & 36 & 06.95 & 23.13 & 20.14 &  \nd  & 0.22 & 0.15 &  \nd &  0.04\\
D33 J013251.2+303944.1 & 01 & 32 & 51.22 & 30 & 39 & 44.10 & 21.63 &  \nd  & 22.44 & 0.14 &  \nd & 0.11 & -0.02\\
D33 J013251.2+303757.4 & 01 & 32 & 51.22 & 30 & 37 & 57.43 & 22.21 & 21.01 &  \nd  & 0.18 & 0.17 &  \nd &  0.08\\
D33 J013251.2+303855.7 & 01 & 32 & 51.25 & 30 & 38 & 55.68 & 22.79 & 21.24 &  \nd  & 0.30 & 0.25 &  \nd & -0.12\\
D33 J013251.2+303947.4 & 01 & 32 & 51.25 & 30 & 39 & 47.45 & 22.91 & 21.56 &  \nd  & 0.28 & 0.23 &  \nd &  0.11\\
D33 J013251.3+303646.9 & 01 & 32 & 51.25 & 30 & 36 & 46.87 & 22.00 & 20.37 &  \nd  & 0.15 & 0.13 &  \nd & -0.09\\
D33 J013251.3+303936.8 & 01 & 32 & 51.27 & 30 & 39 & 36.79 & 22.67 & 21.10 &  \nd  & 0.23 & 0.19 &  \nd &  0.08\\
D33 J013251.3+304013.5 & 01 & 32 & 51.30 & 30 & 40 & 13.51 & 22.31 & 22.10 & 22.35 & 0.19 & 0.34 & 0.09 & -0.05\\
D33 J013251.3+303952.3 & 01 & 32 & 51.30 & 30 & 39 & 52.34 & 22.64 & 21.34 &  \nd  & 0.22 & 0.18 &  \nd &  0.02\\
D33 J013251.3+303913.1 & 01 & 32 & 51.31 & 30 & 39 & 13.11 & 20.88 & 19.28 & 21.86 & 0.06 & 0.05 & 0.03 &  0.14\\
D33 J013251.3+303802.8 & 01 & 32 & 51.32 & 30 & 38 & 02.76 & 22.27 & 20.93 &  \nd  & 0.18 & 0.19 &  \nd & -0.00\\
D33 J013251.3+303719.3 & 01 & 32 & 51.32 & 30 & 37 & 19.27 & 22.70 & 19.79 &  \nd  & 0.23 & 0.08 &  \nd &  0.22\\
D33 J013251.3+303650.6 & 01 & 32 & 51.32 & 30 & 36 & 50.61 & 22.88 & 21.59 &  \nd  & 0.29 & 0.22 &  \nd &  0.04\\
D33 J013251.3+303539.1 & 01 & 32 & 51.33 & 30 & 35 & 39.12 & 21.94 & 21.27 &  \nd  & 0.24 & 0.25 &  \nd &  0.45\\
D33 J013251.3+303842.9 & 01 & 32 & 51.34 & 30 & 38 & 42.87 & 22.89 & 20.45 &  \nd  & 0.28 & 0.17 &  \nd &  0.26\\
D33 J013251.3+303701.1 & 01 & 32 & 51.34 & 30 & 37 & 01.06 & 22.35 & 19.72 &  \nd  & 0.18 & 0.11 &  \nd &  0.04\\
D33 J013251.3+303940.1 & 01 & 32 & 51.35 & 30 & 39 & 40.07 & 23.18 & 20.55 & 23.67 & 0.30 & 0.11 & 0.17 &  0.05\\
D33 J013251.3+303644.0 & 01 & 32 & 51.35 & 30 & 36 & 43.95 & 20.57 & 18.78 &  \nd  & 0.06 & 0.04 &  \nd & -0.04\\
D33 J013251.3+303823.0 & 01 & 32 & 51.35 & 30 & 38 & 23.03 & 20.75 & 21.09 & 20.79 & 0.06 & 0.23 & 0.02 & -0.07\\
D33 J013251.3+304009.5 & 01 & 32 & 51.35 & 30 & 40 & 09.48 & 22.86 &  \nd  & 23.20 & 0.36 &  \nd & 0.19 &  0.56\\
D33 J013251.3+303703.3 & 01 & 32 & 51.35 & 30 & 37 & 03.29 & 23.24 & 21.51 &  \nd  & 0.35 & 0.28 &  \nd &  0.11\\
D33 J013251.4+303814.2 & 01 & 32 & 51.35 & 30 & 38 & 14.21 & 23.15 & 21.93 & 23.66 & 0.35 & 0.31 & 0.25 &  0.10\\
D33 J013251.4+303726.9 & 01 & 32 & 51.35 & 30 & 37 & 26.87 & 21.33 & 20.82 & 21.44 & 0.09 & 0.11 & 0.04 &  0.05\\
\multicolumn{14}{l}{\it Continues in electronic form} \\
\enddata
\end{deluxetable}

\begin{deluxetable}{clcc}
\tabletypesize{\normalsize}
\baselineskip=12pt
\tablecolumns{4}
\tablewidth{0pt}
\tablenum{4}
\tablecaption{Photometry comparisons}
\tablehead{\colhead{Band} & \colhead{$\Delta$~mag} & \colhead{$m_{lim}$} & \colhead{N}}
\startdata
\multicolumn{4}{l}{\it Internal -- overlap regions} \\
V & $-0.003\pm0.003$ & 19.5 &   327 \\
I & $+0.014\pm0.002$ & 19.0 &   357 \\
B & $-0.003\pm0.004$ & 19.5 &   160 \\\hline
\multicolumn{4}{l}{\it \citet{wfm90}} \\
V & $-0.041\pm0.058$ & 20.0 &    26 \\
I & $-0.032\pm0.068$ & 20.0 &    30 \\
B & $-0.024\pm0.108$ & 20.0 &    20 \\\hline
\multicolumn{4}{l}{\it \citet{be01}} \\
V & $-0.031\pm0.047$ & 18.5 &    39 \\
B & $-0.038\pm0.033$ & 18.5 &    31 \\
\enddata
\end{deluxetable}

\begin{deluxetable}{ccccc}
\tabletypesize{\normalsize}
\baselineskip=12pt
\tablecolumns{5}
\tablewidth{0pt}
\tablenum{5}
\tablecaption{Artificial star tests -- Results}
\tablehead{\colhead{Band} & \colhead{Mag.} & \multicolumn{3}{c}{$\Delta$~mag (meas-input)} \\
\colhead{} & \colhead{} & \colhead{median} & \colhead {mean} & \colhead{$\sigma$}}
\startdata
V & 15.5 & -0.004 & -0.007 & 0.019\\
  & 16.5 & -0.009 & -0.016 & 0.025\\
  & 17.5 & -0.016 & -0.026 & 0.040\\
  & 18.5 & -0.028 & -0.034 & 0.053\\
  & 19.5 & -0.041 & -0.052 & 0.092\\
  & 20.5 & -0.062 & -0.081 & 0.158\\\hline
I & 15.5 & -0.014 & -0.020 & 0.025\\
  & 16.5 & -0.025 & -0.032 & 0.040\\
  & 17.5 & -0.039 & -0.050 & 0.066\\
  & 18.5 & -0.063 & -0.076 & 0.084\\
  & 19.5 & -0.076 & -0.089 & 0.146\\
  & 20.5 & -0.081 & -0.128 & 0.248\\\hline
B & 16.5 & -0.006 & -0.012 & 0.021\\
  & 17.5 & -0.013 & -0.020 & 0.034\\
  & 18.5 & -0.019 & -0.030 & 0.048\\
  & 19.5 & -0.029 & -0.039 & 0.076\\
  & 20.5 & -0.039 & -0.055 & 0.132\\
  & 21.5 & -0.046 & -0.068 & 0.225\\
\enddata
\end{deluxetable}
\end{document}